\PassOptionsToPackage{pdftex,final}{graphicx}
\documentclass[a4paper,final]{llncs}

\usepackage[lighttt]{lmodern}
\usepackage{amssymb}
\usepackage[T1]{fontenc}
\usepackage[utf8]{inputenc}
\usepackage{amsmath}
\usepackage{amsfonts}
\usepackage{stmaryrd}
\usepackage[usenames,dvipsnames]{xcolor}
\usepackage{graphicx}
\usepackage{fancyvrb}
\usepackage{longtable}
\usepackage{comment}
\usepackage{xspace}
\usepackage{url}
\usepackage{ifdraft}
\usepackage[final]{listings}
\usepackage{enumerate}
\usepackage{endnotes}
\usepackage[pdftex,bookmarks=true]{hyperref}
\usepackage{wrapfig}
\hypersetup{
  pdfauthor = {Jacek Chrząszcz and Aleksy Schubert},
  pdftitle = {},
  pdfsubject = {D.3.2, D.3.3, D.1.1, D.1.5, F.3.2},
  pdfkeywords = {Java, Object-oriented Programming, Type Systems},
  pdfcreator = {LaTeX with hyperref package},
  pdfproducer = {pdflatex},
  bookmarksdepth=3,
  bookmarks=true,
  bookmarksopen=true,
  bookmarksopenlevel=3,
  final
}

\newcommand{\draft}[1]{\ifdraft{{\color{red} [[[#1]]]}}{} }
\newcommand{\draftmargin}[1]{\ifdraft{\marginpar{\parbox{2cm}{\color{red}\raggedright\small
        #1}}}{\ignorespaces}}

\ifdraft{\DefineVerbatimEnvironment{draftverbatim}{Verbatim}{formatcom=\color{red}}}{\specialcomment{draftverbatim}{}{}\excludecomment{draftverbatim}}

\newcommand{\draddbookmark}[1]{\ifdraft{\phantomsection\addcontentsline{toc}{subsubsection}{#1}}{\ignorespaces}}

\newcommand{\Jafun}{\textsl{Jafun}\xspace}

\newcommand{\ls}{\mathtt{rwr}\xspace}
\newcommand{\lenv}{\mathtt{rd}\xspace}
\newcommand{\bb}{\mathtt{atm}\xspace}

\newcommand{\dsrul}[1]{\hypertarget{srul-#1}{\textrm{(#1)}}} 
\newcommand{\srul}[1]{\hyperlink{srul-#1}{\textrm{(#1)}}} 
\newcommand{\dtrul}[1]{\hypertarget{trul-#1}{\raisebox{0.2ex}{\rm \scriptsize
    (}\textsc{#1}\raisebox{0.2ex}{\rm \scriptsize )}}}

\newcommand{\mmod}{\mu\xspace}

\newcommand{\fmodifier}{\phi\xspace}
\newcommand{\argone}{\mathsf{arg}\xspace}
\newcommand{\argonen}{\mathsf{argn}\xspace}
\newcommand{\args}{\overline{\mathsf{arg}}\xspace}
\newcommand{\argns}{\overline{\mathsf{argn}}\xspace}
\newcommand{\Ex}{\mathsf{Exc}\xspace}
\newcommand{\Exn}{\mathsf{Excn}\xspace}
\newcommand{\Exc}{\overline{\mathsf{Exc}}\xspace}
\newcommand{\Excn}{\overline{\mathsf{Excn}}\xspace}
\newcommand{\fieldref}{\mathsf{fieldref}\xspace}
\newcommand{\varref}{\mathsf{v}\xspace}
\newcommand{\varrefs}{\overline{\varref}}

\newcommand{\jclass}{\mathbf{class}\xspace}
\newcommand{\jext}{\mathbf{ext}\xspace}

\newcommand{\reg}{\emptyset}
\newcommand{\rep}{\mathtt{rep}\xspace}
\newcommand{\throws}{\mathbf{throws}\xspace}
\newcommand{\jlet}{\mathbf{let}\xspace}
\newcommand{\jin}{\mathbf{in}\xspace}
\newcommand{\jif}{\mathbf{if}\xspace}
\newcommand{\jthen}{\mathbf{then}\xspace}
\newcommand{\jelse}{\mathbf{else}\xspace}
\newcommand{\jthrow}{\mathbf{throw}\xspace}
\newcommand{\jtry}{\mathbf{try}\xspace}
\newcommand{\jcatch}{\mathbf{catch}\xspace}
\newcommand{\jnull}{\mathbf{null}\xspace}
\newcommand{\jnew}{\mathbf{new}\xspace}
\newcommand{\jthis}{\textbf{this}\xspace}
\newcommand{\letin}[4]{\jlet\; #1\; #2 = #3\; \jin\; #4\xspace}
\newcommand{\ite}[3]{\jif\; #1\; \jthen\; #2\; \jelse\; #3\xspace}
\newcommand{\newin}[3]{\jnew\; #1\; #2(#3)\xspace}
\newcommand{\throwin}[1]{\jthrow\; #1\xspace}
\newcommand{\tcatch}[4]{\jtry\; \boldsymbol{\{}#1\boldsymbol{\}}\; \jcatch\; (#2\; #3)\; \boldsymbol{\{}#4\boldsymbol{\}}\xspace}

\newcommand{\Cname}{\mathsf{CId}\xspace}
\newcommand{\Lannot}{\mathsf{AMod}\xspace}

\newcommand{\Ident}{\mathsf{Id}\xspace}
\newcommand{\MIdent}{\mathsf{MId}\xspace}
\newcommand{\Expr}{\mathsf{Expr}\xspace}
\newcommand{\BCtxt}{\mathsf{BCtxt}\xspace}
\newcommand{\Stacks}{\mathsf{Stacks}\xspace}

\newcommand{\Loc}{\mathsf{Loc}\xspace}
\newcommand{\Heap}{\mathsf{Heap}\xspace}
\newcommand{\ctxt}{\mathcal{C}\xspace}
\newcommand{\alloc}{\mathsf{alloc}\xspace}
\newcommand{\Prog}{\mathsf{Prog}\xspace}
\newcommand{\body}{\mathsf{body}\xspace}
\newcommand{\classof}{\mathsf{class}\xspace}

\newcommand{\fields}{\mathsf{flds}\xspace}

\newcommand{\paramNames}{\mathsf{parNms}\xspace}

\newcommand{\emptyclass}[1]{\mathsf{empty}_{#1}\xspace}

\newcommand{\npetype}{\texttt{NPE}\xspace}

\newcommand{\objecttype}{\texttt{Object}\xspace}
\newcommand{\npe}{\mathsf{npe}\xspace}

\lstdefinelanguage{Jafun}{
  language=Java,
  basicstyle=\small\ttfamily\upshape,
  keywordstyle=\bfseries\ttfamily\small,
morekeywords={rd,rwr,atm,rep,peer,readonly,let,in,Pure,Fresh,accessible,assignable,mutable,polyread,readable,writable,immutable,isolated},
  escapeinside=||,
}

\lstdefinelanguage{Coq}%
{morekeywords={Variable,Section,Inductive,CoInductive,Fixpoint,CoFixpoint,Declare,%
    Definition,Lemma,Theorem,Axiom,Local,Save,Grammar,Syntax,intro,Eval,comput
e,%
      trivial,Qed,intros,decompose,and,symmetry,admit,simpl,rewrite,Resolve,apply,elim,assumption,%
      left,cut,case,auto,intuition,forall,fun,unfold,exact,right,Hypothesis,patt
ern,destruct,eqn,%
      constructor,Defined,fix,Record,Proof,induction,Hints,exists,let,in,%
      Parameter,split,reflexivity,transitivity,if,then,else,Opaque,%
      Transparent,inversion,inversion_clear,absurd,generalize,Mutual,match,of,en
d,Analyze,struct,Ltac,%
      with,by,as,Mutual,Rewall,Set,Prop,Type,%
      Module,Import,End,Time,Require,Open,Scope,Export,repeat,Extraction,Notation,return,%
      AutoRewrite,Functional,Scheme,params,refine,using,discriminate,try,eapply,assert,case_eq,context},%
   sensitive,%
   keywordstyle=\bfseries,
   basicstyle=\small\slshape,
   morecomment=[n]{(*}{*)}, 
   literate={:=}{{$:=$}}1 
{|}{{$\!\vert$}}1 
{|-}{{$\vdash$}}1 
{<-}{{$\!\leftarrow\!$}}1 
{->}{{$\;\rightarrow\;$}}1 
{=>}{{$\Rightarrow\;$}}1 
{/\\}{{$\land\;$}}1 
{\\/}{{$\lor\;$}}1 
{::}{\,{::}\;}1 
{<->}{{$\leftrightarrow$}}1
{[}{{$[$}}1 
{_}{{\tiny\_}}1
{_[}{{{\tiny\_}$[\,$}}1 
{[[}{{$\,[\,[\,$}}1 
{_[[_}{{{\tiny\_}$[\,[${\tiny\_}}}2 
{]}{{$]$}}1
{]_}{{$\;]${\tiny\_}}}1
{]]}{{$\,]\,]$}}2
{]]_}{{$\,]\,]${\tiny\_}}}2 
{_]]_}{{{\tiny\_}$]\,]${\tiny\_\ }}}2
{__}{{\tiny\_\_}}2,%
   morestring=[d]",
   showstringspaces=false,
   escapeinside=!!,
  }

\lstnewenvironment{lstcoq}{\lstset{language=Coq}}{}
\newcommand{\coqinl}[1]{\lstinline[language=Coq]{#1}}

\lstnewenvironment{lstjafun}{\lstset{language=Jafun}}{}
\newcommand{\jainl}[1]{\lstinline[language=Jafun]{#1}}

\begin{document}

\title{Formalisation of a frame stack semantics\\ 
  for a~Java-like language\thanks{This work was partially supported by the Polish NCN
  grant 2013/11/B/ST6/01381.}}

\author{Aleksy Schubert and Jacek Chrząszcz}
\institute{Institute of Informatics, University of Warsaw\\
   ul. S. Banacha 2, 02--097 Warsaw, Poland\\
\email{\tt [alx,chrzaszcz]@mimuw.edu.pl}
}

\maketitle

\begin{abstract}
  We present a Coq formalisation of the small-step operational
  semantics of \Jafun, a small Java-like language with classes. This
  format of semantics makes it possible to naturally specify and prove
  invariants that should hold at each computation step.  In contrast
  to the Featherweight Java approach the semantics explicitly
  manipulates frame stack of method calls. Thanks to that one can
  express properties of computation that depend on execution of
  particular methods. 

  On the basis of the semantics, we developed a type system that makes
  it possible to delineate a notion of a compound value and classify
  certain methods as extensional functions operating on them. In our
  formalisation we make a mechanised proof that the operational
  semantics for the untyped version of the semantics agrees with the
  one for the typed one. We discuss different methods to make such
  formalisation effort and provide experiments that substantiate it.
\end{abstract}

\section{Introduction}

The small-step semantics \cite{Plotkin81} can serve as a framework in
which interesting invariant properties of computations are naturally
expressed. The primary reason for this is that by definition
the property should hold at \emph{each computation step}. Both in
big-step semantics \cite{Kahn87} and in denotational semantics
\cite{ScottS71} the main focus is on the resulting
value and the syntactical structure of the program expression at hand
while the intermediate computational steps become hidden.

In design of small-step semantics one can decide to take the
Featherweight Java (FJ) \cite{IgarashiPW01} approach, in which a
method body is directly expanded in place of its call. In this
approach, one gets a small formal machinery that is simpler to work
with. Still, this formal language model is too simple to relate
certain desired properties of interest. A richer model in which
the method call stack is directly represented makes it possible to
deal with the following cases.
\begin{itemize}
\item It is possible to directly represent the management of static
  scopes, which is vital for escape analysis \cite{StilkerichLEBS17}
  and simplifies some optimisation analyses.  
\item It is possible to express properties of evolving call stacks
  (e.g.\ that the call stack belongs to a~particular regular language,
  a property that occurs in many security related specifications
  \cite{Chang06}).
\item It is possible to directly express strong computational
  invariants that 
  require management of access scopes at entry and exit of a method,
  these include immutability \cite{HaackPollSchubert07} or
  functionality \cite{ChrzaszczS17}.
\end{itemize}
These advantages are also recognised among the authors of notable
formalisations (e.g.\ recently CompCert turns largely to small-step
semantics \cite{Leroy12}) although other approaches, e.g.\
co-inductive one as in \cite{LeroyG09,MoorePR18}, may give similar
results.

The use of method frame stack was vital for our paper and pencil
soundness proof of a type system that makes it possible to delineate
the notion of a compound value in a Java-like language and define
extensional functions that operate on such values
\cite{ChrzaszczS17}. Since there was no attempt to make a mechanised
formalisation of a Java-like language small-step semantics with
evaluation based upon method frame stack we decided to develop one and
take our type system as a test bed for various approaches to
formalisation and discuss their consequences.

As a consequence of these efforts we obtained a formalised semantics
of a~Java-like language \Jafun\footnote{The formalisation is available
  from \url{http://www.mimuw.edu.pl/~alx/jafun.tgz}} with
\begin{itemize}
\item a hierarchy of classes and related subtyping relation;
\item small-step reduction relation defined in terms of method frame
  stacks;
\item an example type system of the language that captures the notion
  of value and extensional functions that transform such values;
\item a Church-style version of the type system together with proofs
  that the operational small-step semantics of the Church-style
  version agrees with the original, untyped semantics.
\end{itemize}
On the basis of these formalisation artefacts, we discuss various
design decisions and their consequences for development of such
semantics, which can be useful in other formalisation efforts. In
particular, we stress the following points.
\begin{itemize}
\item The natural way to define small-step semantics relation in Coq
  may result in duplication of cases and cause excessive proving
  efforts. We propose a methodology to transform the natural
  definition into a sparing one and demonstrate savings it brings in
  proof development.
\item Various coherence proofs, e.g.\ soundness and completeness of the
  typed reduction with regard to the untyped one, require case
  analysis with large number of cases. Any attempt at
  automatising of the case analysis requires the
  person who develops the proof to frequently recognise which case is
  currently analysed. We propose a method that makes this task 
  significantly easier.
\item When automatic case analysis is employed, different
  strategies of discharging the cases can be used. We discuss the
  advantages and disadvantages of two approaches to case analysis. In
  the first one, we destruct all the available case distinctions and
  discharge cases where no longer case analysis is possible. In the
  second one, we destruct case distinctions in only one definition and
  apply its results in all the remaining ones. It turns out that
  although the first approach is more general, the second one results
  in shorter proofs.
\end{itemize}

This paper is organised as
follows. Section~\ref{sec:syntax-and-semantics} introduces our
language.  In Section~\ref{sec:typed}, we present
the Church-style typed version of our semantics while in
Section~\ref{sec:experiments} we discuss the advantages of different
ways its coherence proofs can be done.  At the end in
Section~\ref{sec:related-works} we give an account of works that are
related to our efforts and at last in Section~\ref{sec:conclusions} we
summarise our results.


\section{Syntax and semantics of \Jafun}
\label{sec:syntax-and-semantics}

\expandafter\ifx\csname SourceFile\endcsname\relax\else\SourceFile{semantics-f.ltx}\fi
\paragraph{Main features of \Jafun.}
%


\begin{figure}[tbp]
\begin{lstjafun}
class DList {  rep DList prev;  Data val;  rep DList next;|\smallskip|                        
  rwr DList rd copy() {  return this.appRec(null);  }    |\smallskip|
  rwr DList rd appRec(rwr DList newPrev) {        
    DList newThis = new rwr DList(newPrev, val, null); 
    if (newPrev != null) { newPrev.next = newThis; }
    if (next != null) { newThis.next = next.appRec(newThis); }
    return newThis;  }     |\smallskip|
  rwr DList atm singleton(atm Data v) {  
    return new rwr DList(null, v, null);  }     
}                                          
\end{lstjafun}
\caption{An example of \Jafun annotations: a doubly linked list (in Java syntax)}
\label{fig:dlist}
\end{figure}

Fig.~\ref{fig:dlist} presents an example implementation of doubly
linked list with \Jafun annotations, written in Java syntax for
clarity.
%
%
%
The $\rep$ annotations for fields (here: \jainl{prev} and \jainl{next})
serve to establish a notion of compound value in \Jafun. When a field is
marked with $\rep$ the reference in the field points to
further representation of the value.
If a field is not marked (here: \jainl{val}), 
the reference stored in the field is part of the
current value representation, but the object it points to is not.
Thanks to these annotations, a one element list containing a given
object can be considered the same compound value at two points of a
program execution even if the state of the contained object changes
between the two points. Moreover, any single element lists can be 
considered equal, even in different runs of the program, when one 
considers only single argument list operations. 

The goal of the \Jafun type system is to establish that suitably
annotated methods are extensional functions, i.e., always yield
equal results when applied to equal values (e.g. at two points of a
program execution) and do not change the state of pre-existing
objects (for precise definition of extensional functions and compound value equality in an object-oriented context, we refer the reader to~\cite{ChrzaszczS17}). 
Such methods however can imperatively modify the state of newly
created objects, also using auxiliary methods. 
This programming style is very flexible as demonstrated by Okasaki
\cite{Okasaki99}, but the modifications performed by
the auxiliary methods have to be strictly controlled and that is what
access mode annotations $\ls, \lenv, \bb$ are for.

In the example from Fig.~\ref{fig:dlist}, the \jainl{copy} method
is annotated as a function: its only argument ($\jthis$) is annotated
as $\lenv$, i.e. read-only. The other annotation $\ls$ (for read-write)
is the annotation of the result of the method, meaning in fact that
the result is a freshly allocated object. The auxiliary method
\jainl{appRec} has similar annotations and its argument
\jainl{newPrev} is also annotated as read-write. Its annotations
mean that \jainl{appRec} is not a function, as it can modify its
argument \jainl{newPrev}, but $\jthis$ will remain unmodified
(unless their representation is shared). Indeed, if
\jainl{newPrev} points to the last cell of the copy of the
beginning of the current list (until \jainl{prev}), the method
\jainl{appRec} will correctly append the copy of rest of the 
current list to this cell, without modifying the current list. In the
end it is clear that the method \jainl{copy} is indeed a function,
returning an identical fresh copy of the original unmodified list.


In order to make sure the result of functions do not depend on
internal elements of non-rep fields of objects, as they are not part
of value representations, such references should be followed neither
for reading nor writing which brings to the system a kind of sealed
references which need however to be passed around also at the
interprocedural level. To mark such references we use the $\bb$
annotation (for atomic), which are used in the \jainl{singleton}
function in Fig.~\ref{fig:dlist} (which ignores its $\jthis$ argument).

To complete the picture, we remark that parts of objects marked with
$\lenv$ (or $\bb$) can be modified (or read and modified), but this
can only be done through a~different variable with suitable access
mode that gives permission to write (or read and write) to the
representation. 

The complete syntax of \Jafun is given in~Fig.~\ref{fig:syntax}. Apart
from annotations, its main differences wrt.\ Java consist in
replacing all instructions with expressions, introducing $\jlet$ expressions, restricting expressions in many
positions to identifiers (but thanks to $\jlet$ that does not restrict expressiveness) and
replacing sequencing with a semi-colon by sequencing with $\jlet$. To
further simplify the language we do not consider visibility
annotations (everything is \texttt{public}) and assume that every
class has a single built-in ``assign to all fields'' constructor (that
assumption is used in Fig.~\ref{fig:dlist}). The last two elements in Fig.~\ref{fig:syntax} are needed for the reduction semantics and are explained in Sect.~\ref{sec:semantics}. 

More details 
of the language
can be found in~\cite{ChrzaszczS17}.

\begin{figure}[!t]
  \vspace{-0.7em}
\draddbookmark{SYNTAX}%
\centering
\begin{displaymath}%
 \begin{array}{@{}r@{\,}l@{\;\;}c@{\;\;}l@{}}
    \Prog\owns
    & \mathbf{C}          & ::= & \jclass\;
                       C_1\; \jext\; C_2\; \boldsymbol\{ \overline{\mathbf{F}}~ \overline{\mathbf{M}} \boldsymbol\}\\
    \Cname\owns
    & C          & ::= & \langle \textit{identifier}\rangle \quad\textit{(class name)}\\
    \Lannot\owns
    & \mmod & ::= & \ls\;|\; \lenv\;|\; \bb\quad\;\;\,
                    \fmodifier ::= \rep\;|\; \reg \\
    & \mathbf{F}         & ::= & \fmodifier\; C\; x\\
    \Ident\owns
    & x          & ::= & \langle \textit{identifier}\rangle \quad\textit{(variable/field name)}\\[1ex]
    & \argone    & ::= & \mmod\; C\; x \qquad\qquad\quad
      \argonen   ~\, ::=  \reg\; C\; x\\
    & \Ex        & ::= & \mmod\; C\qquad\qquad\qquad
      \Exn        ::=  \reg\; C\\[1ex]
    & \mathbf{M}         & ::= & \mmod\; C\; \mmod\; m(\args)\; \throws\; \Exc\;
                         \boldsymbol{\{}E\boldsymbol{\}}\;|\; \reg\; C\; \reg\; m(\argns)\; \throws\; \Excn\; \boldsymbol{\{}E\boldsymbol{\}}\; \\
    \MIdent\owns
    & m          & ::= & \langle \textit{identifier}\rangle \quad\textit{(method name)}\\
    \Expr\owns 
    & E          & ::= & \newin{\mmod}{C}{\varrefs}\;|\;                          \letin{C}{x}{E_1}{E_2}\;|\; \fieldref\;|\; \fieldref = \varref\;|\; \\
    &            &     & \ite{\varref_1 == \varref_2}{E_3}{E_4}\;\;|\;                          \varref.m( \varrefs)\;|\; 
                         \varref\;\;|\;                          \throwin{\varref}\;|\; \\
    &            &     & \tcatch{E_1}{\mmod\; C}{x}{E_2} \\[1ex]
    \multicolumn{2}{r}{\varref\,}    & ::= & x\;|\; \jthis\;|\; \jnull \qquad
    \fieldref\,   ::=  \varref.x 
    \\[1ex]                                         
      & A      & ::= & C\;|\; \emptyset\\
    \BCtxt\owns
      & \ctxt     & ::= &
                     \llbracket ~\rrbracket _A\;|\; \letin{C}{x}{\ctxt}{E}\;\;|\;                        \tcatch{\ctxt}{\mmod\; C}{x}{E}\end{array}
\end{displaymath}%
\caption{Abstract syntax of \Jafun}%
\label{fig:syntax}%
\end{figure} 

\paragraph{Abstract syntax and its formalisation.} 
\label{sec:abstract-syntax}

The syntax of \Jafun (see Fig.~\ref{fig:syntax})
is reflected in our formalisation as closely as possible by inductive
types. For instance class declarations are defined as 
\begin{lstcoq}
Inductive JFClassDeclaration : Set :=
| JFCDecl (cn:JFClassName) (ex:option JFClassName)
          (fields:list JFFieldDeclaration)
          (methods:list JFMethodDeclaration).
\end{lstcoq}
which follows the structure of the corresponding grammar rule.
We use the \coqinl{option} type here to
represent the variant of class declaration which is not extended 
(possible only for \texttt{Object} in well formed programs).
We use \coqinl{option} type systematically to convey
that some element of syntax may be missing.

A program in \Jafun is a list $ \overline{\mathbf{C}}$ of \emph{class declarations}
with unique names that contains two predefined classes
\objecttype and \npetype (for \texttt{NullPointerException}). In our
semantics, we represent programs as 
\begin{lstcoq}
Definition JFProgram : Set := list JFClassDeclaration.
\end{lstcoq}
This choice reflects the situation before the program is loaded into
the memory and enables the study of the basic properties that are
important for proper execution of the language and are enforced by the
loading process.  Still, this approach requires us to formulate and
maintain in proofs certain well-formedness conditions (gathered in the
Coq predicate \coqinl{Well_formed_program CC}), since e.g. in
post-loading view a program cannot contain duplicate class
declarations.

\subsection{Overview of mechanised \Jafun semantics}\label{sec:semantics}

\begin{figure*}[p]
\begin{center}
\draddbookmark{SEMANTIC RULES}
\small
\begin{displaymath}%
  \begin{array}{l@{\;}l}
  \overline{\mathbf{C}},\ h,\,  \ctxt_1\llbracket E_1\rrbracket _{A_1}::\cdots:: & \ctxt_n\llbracket E_n\rrbracket _{A_n}\;\;\to\;\;\;
      h', \ctxt'_1\llbracket E_1\rrbracket _{A'_1}::\cdots:: \ctxt'_m\llbracket E_m\rrbracket _{A'_m}.
  \end{array}\end{displaymath}
\vspace{1ex}
\begin{tabular}{@{}ll@{}}
\dsrul{newk}
&

$%
 \overline{\mathbf{C}}, h,  \overline{\ctxt}:: \ctxt\llbracket \newin{\mmod}{C}{l_1,\ldots,l_k}\rrbracket _\emptyset \rightarrow h'', \overline{\ctxt}:: \ctxt\llbracket l_0\rrbracket _\emptyset$
\\
\multicolumn{2}{l}{\qquad where
$%
\alloc(h, \overline{\ctxt}, C)=(l_0, h'),\;
\fields(C) = x_1, \dots, x_k,\;$}\\
\multicolumn{2}{l}{
\qquad $%
o=\emptyclass{C}\{x_1 \mapsto  l_1,\ldots,x_k \mapsto  l_k\}, \;
h'' = h'\{l_0 \mapsto  o\}$}\\[2ex]
\dsrul{letin}
&
$%
 \overline{\mathbf{C}}, h, \overline{\ctxt}:: \ctxt\llbracket \letin{C}{x}{E_1}{E_2}\rrbracket _\emptyset\rightarrow h, \overline{\ctxt}:: \ctxt[\letin{C}{x}{\llbracket E_1\rrbracket _\emptyset}{E_2}]$
\\
\dsrul{letgo}
&
$%
 \overline{\mathbf{C}}, h, \overline{\ctxt}:: \ctxt[\letin{C}{x}{\llbracket l\rrbracket _\emptyset}{E}]\rightarrow h, \overline{\ctxt}:: \ctxt\llbracket E\{l/x\}\rrbracket _\emptyset$
\\[2ex]
\dsrul{ifeq} 
&
$%
 \overline{\mathbf{C}}, h, \overline{\ctxt}:: \ctxt\llbracket \ite{ l_0 == l_1}{E_1}{E_2}\rrbracket _\emptyset\! \rightarrow \!h, \overline{\ctxt}:: \ctxt\llbracket E_1\rrbracket _\emptyset$
\quad where
$l_0=l_1$
\\
\dsrul{ifneq}
&
$%
 \overline{\mathbf{C}}, h, \overline{\ctxt}:: \ctxt\llbracket \ite{ l_0 == l_1}{E_1}{E_2}\rrbracket _\emptyset\! \rightarrow \!h, \overline{\ctxt}:: \ctxt\llbracket E_2\rrbracket _\emptyset$
\quad where
$l_0\not=l_1$
\\[2ex]
\dsrul{mthdnpe}
&
$%
 \overline{\mathbf{C}}, h, \overline{\ctxt}:: \ctxt\llbracket \jnull.m( \overline{l})\rrbracket _\emptyset \rightarrow h, \overline{\ctxt}:: \ctxt\llbracket \npe\rrbracket _\npetype$
\\
\dsrul{mthd}
&
$%
 \overline{\mathbf{C}}, h, \overline{\ctxt}:: \ctxt\llbracket l.m(\overline{l})\rrbracket _\emptyset \rightarrow h, \overline{\ctxt}:: \ctxt\llbracket l.m(\overline{l})\rrbracket _\emptyset::\llbracket E\rrbracket _\emptyset$
\\
\multicolumn{2}{l}{\qquad where
$%
 \classof(h, l)=D, \body(D, m)=E_0, E = E_0\{l/\jthis, \overline{l}/\paramNames(D, m)\} $
}
\\[1ex]
\dsrul{mthdret}
&
$%
 \overline{\mathbf{C}}, h, \overline{\ctxt}:: \ctxt\llbracket l.m(\overline{l})\rrbracket _\emptyset::\llbracket l'\rrbracket _\emptyset \rightarrow h, \overline{\ctxt}:: \ctxt\llbracket l'\rrbracket _\emptyset$
\\[2ex]
\dsrul{assignnpe}
&
$%
 \overline{\mathbf{C}}, h, \overline{\ctxt}:: \ctxt\llbracket \jnull.x = l\rrbracket _\emptyset \rightarrow h, \overline{\ctxt}:: \ctxt\llbracket \npe\rrbracket _\npetype$
\\
\dsrul{assignev}
&
$%
 \overline{\mathbf{C}}, h, \overline{\ctxt}:: \ctxt\llbracket l_1.x = l\rrbracket _\emptyset \rightarrow h', \overline{\ctxt}:: \ctxt\llbracket l\rrbracket _\emptyset$
\\
\multicolumn{2}{l}{
\qquad where
$ l_1 \neq \jnull, o = h(l_1)\{x \mapsto  l\}, h' = h\{l_1 \mapsto  o\} $
}
\\[2ex]
\dsrul{varnpe}
&
$%
 \overline{\mathbf{C}}, h, \overline{\ctxt}:: \ctxt\llbracket \jnull.x\rrbracket _\emptyset \rightarrow h, \overline{\ctxt}:: \ctxt\llbracket \npe\rrbracket _\npetype$
\\
\dsrul{var}
&
$%
 \overline{\mathbf{C}}, h, \overline{\ctxt}:: \ctxt\llbracket l.x\rrbracket _\emptyset \rightarrow h, \overline{\ctxt}:: \ctxt\llbracket l'\rrbracket _\emptyset$
\qquad where
$ l\not=\jnull,    l'=h(l)(x) $
\\[2ex]
\dsrul{thrownull}
&
$%
\overline{\mathbf{C}}, h, \overline{\ctxt}:: \ctxt\llbracket \throwin{\jnull}\rrbracket _\emptyset \rightarrow h, \overline{\ctxt}:: \ctxt\llbracket \npe\rrbracket _{\npetype}$
\\[2ex]
\dsrul{throw}
&
$%
 \overline{\mathbf{C}}, h, \overline{\ctxt}:: \ctxt\llbracket \throwin{l}\rrbracket _\emptyset \rightarrow h, \overline{\ctxt}:: \ctxt\llbracket l\rrbracket _{D}$
\qquad where
$l\not=\jnull,    \classof(h, l) = D$
\\[2ex]
\dsrul{ctchin}
&
$%
 \overline{\mathbf{C}}, h, \overline{\ctxt}:: \ctxt\llbracket \tcatch{E_1}{\mmod\; C}{x}{E_2}\rrbracket _\emptyset \rightarrow $\\

&
$ \phantom{\overline{\mathbf{C}},\;} 
h, \overline{\ctxt}:: \ctxt[\tcatch{\llbracket E_1\rrbracket _\emptyset}{\mmod\; C}{x}{E_2}] $
\\
\dsrul{ctchnrml}
&
$%
 \overline{\mathbf{C}}, h, \overline{\ctxt}:: \ctxt[\tcatch{\llbracket l\rrbracket _\emptyset}{\mmod\; C}{x}{E_2}] \rightarrow h, \overline{\ctxt}:: \ctxt\llbracket l\rrbracket _\emptyset$
\\
\dsrul{ctchexok}
&
$%
 \overline{\mathbf{C}}, h, \overline{\ctxt}:: \ctxt[\tcatch{\llbracket l\rrbracket _{C'}}{\mmod\; C}{x}{E_2}] \rightarrow h, \overline{\ctxt}:: \ctxt\llbracket E_2'\rrbracket _\emptyset$ \\
\multicolumn{2}{l}{
\qquad where
$%
E_2' = E_2\{l/x\}, C' \leq:  C$ }
\\[2ex]
\dsrul{letex}
&
$%
 \overline{\mathbf{C}}, h, \overline{\ctxt}:: \ctxt[\letin{C}{x}{\llbracket l\rrbracket _{C'}}{E}]\rightarrow h, \overline{\ctxt}:: \ctxt\llbracket l\rrbracket _{C'}$
\qquad where
$C'\not=\emptyset$  
\\
\dsrul{methodex}
&
$%
 \overline{\mathbf{C}}, h, \overline{\ctxt}:: \ctxt\llbracket l.m(\overline{l})\rrbracket _\emptyset::\llbracket l'\rrbracket _C \rightarrow h, \overline{\ctxt}:: \ctxt\llbracket l'\rrbracket _C$
\qquad where
$C\not=\emptyset$  
\\
\dsrul{ctchexnok}
&
$%
 \overline{\mathbf{C}}, h, \overline{\ctxt}:: \ctxt[\tcatch{\llbracket l\rrbracket _{C'}}{\mmod\; C}{x}{E_2}] \rightarrow h, \overline{\ctxt}:: \ctxt\llbracket l\rrbracket _{C'}$\\
\multicolumn{2}{l}{%
\qquad where
$%
C'\not=\emptyset, C' \not\!\leq:  C$  
}
\\[3ex]
\end{tabular}
\end{center}

We assume here that
$ l, l', l_0, l_1\in\Loc, \overline{l}\in\Loc^* $,
$h, h'\in\Heap$,
$ \overline{\mathbf{C}}\in \Prog$,
$ \overline{\ctxt}\in \Stacks$,
$ \ctxt, \ctxt_1, \dots , \ctxt_n \in \BCtxt$,
$C, D\in \Cname$,
$A_1, \ldots, A_n \in \Cname \cup \{\emptyset\}$,
$m\in \MIdent$,
$x\in \Ident$, 
$E, E_1, \ldots, E_n\in\Expr$, and
$\alloc:\Heap\times\Prog\times\Cname \to \Loc\times\Heap$. 

\caption{Semantic reduction relation of \Jafun}
\label{fig:semantics}
\end{figure*}

\paragraph{Reduction relation.}

The small step semantics of \Jafun is defined with a reduction relation
$ \rightarrow  $ presented in Fig.~\ref{fig:semantics}. 
The relation is defined for a fixed program $ \overline{\mathbf{C}}$ and connects pairs:
heap, frame stack. 
The general form of the relation is given at the top of the figure.
A \emph{frame stack} $%
 \ctxt_1\llbracket E_1\rrbracket _{A_1}::\cdots:: \ctxt_n\llbracket E_n\rrbracket _{A_n}$, or \( \overline{\ctxt} \) for short, is, roughly speaking, a sequence
of \Jafun expressions in which the current
execution point (redex) is marked with a~special (unary) symbol
$\llbracket \,\rrbracket _A $. The subscript $A$ determines here if the execution is
normal ($A=\emptyset$) or exceptional $ A\in \overline{\mathbf{C}}$. 
Each expression on the stack is divided into an \emph{evaluation context}
(the ``outer layer''), denoted by \( \ctxt_i \in \BCtxt\) (in Fig.~\ref{fig:syntax}), and the redex
$E_i$. Since the evaluation context is already partially computed, its
syntax essentially comes from a much restricted subset $\BCtxt$ of
$\Expr$. Naturally, in a stack frame as above, the redexes
$E_1,\ldots, E_{n-1}$ are method calls $o.m(\overline{v})$, 
and $A_1=\ldots=A_{n-1}=\emptyset$, since a
pending exception is actively dispatched only in the topmost (i.e.~rightmost) frame.

The reduction rules either define the order of execution (basically,
call-by-value from left to right) by specifying how the focus moves
within an context expression (as in~\srul{letin}), or define the meaning of
syntactic constructions (as in~\srul{letgo}).

\paragraph{Formalisation.}
We tried hard to make our Coq formalisation \coqinl{red} of the reduction relation
as visually close to its paper counterpart in Fig.\ref{fig:semantics} as possible. 
Let us introduce its elements.
 
Heaps are defined as maps (suitable instance of a standard library
functor from the FSets collection) from natural numbers to objects
\begin{lstcoq}
Definition Heap : Type := NatMap.t Obj.   
\end{lstcoq}
Objects are pairs that consist of a map from field names (identifiers) to
locations, and a class name
\begin{lstcoq}
Definition RawObj := JFXIdMap.t Loc.
Definition Obj : Type := (RawObj * JFClassName)%type.
\end{lstcoq}
All name and identifier types (\coqinl{JFClassName},
\coqinl{JFXId} etc.) are implemented by standard library ascii
strings.
Locations \coqinl{Loc} are either \coqinl{null} or
natural numbers. Note that this definition of heaps requires us to
define and maintain a heap coherence property which says that keys in
the above mentioned maps agree with field names in the declared class
(expressed as Coq predicate \coqinl{type_correct_heap CC h}).




In our formalisation, we represent the reduction relation as a partial function:
\begin{lstcoq}
Definition red (CC:JFProgram) :
  Heap * FrameStack -> option(Heap * FrameStack) := 
fun '(h, tfs) => match st with    ... !\smallskip!
| (* letin *)
  (Ctx[[JFLet C x E1 E2]]_None) :: Cc  =>  Some (h, 
     (Ctx _[(JFCtxLet C x __ E2) _[[_ E1 _]]_None]_ ) :: Cc)   ... !\smallskip!
| (* mthdret *)
  (nil[[JFVal1(JFVLoc l)]]_None) :: (Ctx[[JFInvoke!{\tiny \ \_ \ \_}!]]_None) :: Cc =>
     Some (h, (Ctx[[ (JFVal1 (JFVLoc l)) ]]_None) :: Cc)   ... !\smallskip!
| _ => None
end.
\end{lstcoq}
The type \coqinl{FrameStack} used above is a list of
\coqinl{Frame}s, each of which is a triple denoted by 
\coqinl{Ctx[[E]]_A}, 
consisting
of a context~\coqinl{Ctx}, an expression
\coqinl{E},
and an execution mode \coqinl{A}, where
\coqinl{None} represents a~normal execution, and
\coqinl{Some D} an exceptional state in which an
exception of class \coqinl{D} is being dispatched.
To ease the manipulation of
contexts (corresponding to $\BCtxt$ in Fig.~\ref{fig:syntax}),
\coqinl{Ctx} is a list of \coqinl{JFCtxLet} or \coqinl{JFCtxTry}
context-elements, where \coqinl{__} (a notation for the constructor \coqinl{tt} of trivial type
\coqinl{unit}) represents a placeholder for deeper
context-expression. The notation \coqinl{Ctx _[Ct _[[E]]_A]_}
used above 
translates to a frame \coqinl{(Ct::Ctx)[[E]]_A}, which represents a context \coqinl{Ctx} nesting \coqinl{Ct} with the redex \coqinl{E} and execution mode~\coqinl{A}.
Thanks to these Coq notations we obtain clear optical correspondence
between the Coq and the paper versions of the semantic rules while
having the comfort of working with 
list operations on our (single-hole) contexts. Apart from some more
noise on the Coq side, the main optical difference between the two
presentations is the order of frames on the stack, but this is only
relevant for a few rules which consider more than one frame at a time.


\section{Type system and typed semantics}
\label{sec:typed}

\expandafter\ifx\csname SourceFile\endcsname\relax\else\SourceFile{typing-f.ltx}\fi
Apart from the semantics 
we define also a type system
for our language. We sketch its structure here since its full
account is presented in~\cite{ChrzaszczS17}. The type system attributes not only a class,
but also an access mode to expressions, specifying what kind of access
to the object at hand is possible in the current context. An example
rule, for the let-expression, is the following:
\begin{displaymath}%
\frac{\begin{array}{c}~ C, m;\; \Xi;\; \Gamma_1 \vdash E_1: \langle C_1, \mmod_1\rangle \\
~ C, m;\; \Xi;\; \Gamma_1, x: \langle C_1, \mmod_1\rangle  \vdash E_2: \langle C_2, \mmod_2\rangle \end{array}}{ C, m;\; \Xi;\; \Gamma_1 \vdash \letin{C_1}{x}{E_1}{E_2}: \langle C_2, \mmod_2\rangle }\quad\textit{\dtrul{let}}\end{displaymath}
The typing relation is specified in Coq as an inductive predicate:
\nopagebreak
\begin{lstcoq}
Inductive types: JFExEnv -> JFEnv -> JFExpr -> JFACId -> Prop := ...
\end{lstcoq}
It expresses that a judgement of the form
$ C, m;\; \Xi;\; \Gamma \vdash E: \tau $ %
holds. The missing arguments for $C,m$ come from 
the~surrounding Coq section 
%
containing a~class and a method declaration 
inside which the typing is supposed to hold.  The correct typing
requires us also to maintain the list of exceptions legal at the
current point together with their access modes (variable $\Xi$
represented in type \coqinl{JFExEnv}), a regular environment
assigning classes and access modes to variables ($\Gamma$
represented in type \coqinl{JFEnv}), the expression at hand
($E$ in type \coqinl{JFExpr}) and the pair consisting of
a~class and an access mode that are assigned to the expression
($\tau$ in type \coqinl{JFACId}). The constructors of the
relation are in one-to-one correspondence with the typing
rules.

From the typing system and the semantics we produce a type annotated
semantics akin to Church versions of type systems in
$\lambda$-calculi. The idea of such a Church version is that the
expression under consideration contains full information concerning
the typing rule that is used to derive its type.
Since our semantics and type system are complex, we observed that any
attempt at 
a~mechanised consistency proof for our type system would be extremely
difficult to get through without a definition of an explicitly typed
version of the frame stack. Therefore, we augment all
frames in the stack with the typing information and state
explicitly many invariants that hold on legal heaps and frame stacks
during the correct execution of the semantics.

The basic structure used in the typed semantics is 
type annotated frame:

\begin{lstcoq}
Record TypedFrame := TFR { 
  TFRcdecl: JFClassDeclaration; TFRmdecl: JFMethodDeclaration;
  TFRXi: JFExEnv;               TFRGamma: JFEnv;
  TFRfr: Frame;                 TFRAcid: JFACId }.
\end{lstcoq}

\noindent
Given that a frame \coqinl{TFRfr} is a triple \coqinl{Ctx[[E]]_A},
the purpose of a typed frame is basically 
that the following typing judgement should be derivable:\\[1.5ex]
\coqinl{       types TFRcdecl TFRmdecl TFRXi TFRGamma (Ctx[[E]]) TFRAcid}\\[-1ex]

\noindent
This, together with some other obvious requirements (e.g. that the
\coqinl{TFRcdecl} class is part of the program) is formalized in
the property called \coqinl{DerivableTFR}.

Next, we state a number of properties that each
frame on the correct typed frame stack should satisfy with respect to the
heap. They are listed in the definition of
\coqinl{oneTFRConsistency} that conveys the following properties:
\begin{itemize}
\item the environment \coqinl{Gamma} from the typed frame
  should contain unique non-null locations,
\item their types declared in \coqinl{Gamma}  are
  supertypes of their types on the heap,
\item a reference for the null pointer exception is present in the
  context.
\end{itemize}

The properties binding every two adjacent typed frames in the stack
are formalised in the inductive definition \coqinl{isTFSind}, these
are in particular
\begin{itemize}
\item each internal frame of the stack represents a method call expression,
  in which method parameters are locations,
\item the method call is on an object that resides in the heap,
\item the class, method and exception context of the subsequent frame
  are respectively the class from which the method is called, the
  method itself, and exceptions allowed by the method declaration,
\item the return type of the method agrees with the target type in the
  next frame.
\draft{Jak mętne to wywalamy :)}
\end{itemize}

The last auxiliary definition in the sequence is \coqinl{isTFS} which
adds to the former two the requirement that if the top frame is in an
exceptional execution mode, then the evaluated expression should be a
valid location corresponding to the class of exception that is being
dispatched. In the very end, \coqinl{DerivableTFS} combines \coqinl{isTFS} with the requirement that each typed frame on the stack is indeed derivable (\coqinl{DerivableTFR}).

\draftmargin{W sumie do definicji next\_costamcostam nie są zupełnie potrzebne te własności, a już zupełnie TFS... Dopiero w lemacie red -> tred}

Now we are ready to define a proper typed semantics:
\begin{lstcoq}
Definition typed_red : 
  Heap * TFSsupport  ->  option (Heap * TFSsupport) := ...
\end{lstcoq}
It is defined in a context with a current program and where
\coqinl{TFSsupport} is a~stack of typed frames. This
function extends the output of the normal semantic function
\coqinl{red} with typing information. For example, the rule for method call looks as follows: 
\begin{lstcoq}
|(* mthd *)
 Ctx[[JFInvoke (JFVLoc (JFLoc n)) m vs]]_None =>
  let D0op := getClassName h n in
  match D0op with | None => None | Some D0 =>
  match getInvokeBody CC D0op n m vs h Ctx (FSofTFS Cc) with 
  |None => None 
  |Some (h', []) => None
  |Some (h', fr :: _) =>
  match find_class CC D0 with None =>None | Some cdecl =>
  match methodLookup CC D0 m with None =>None | Some mdecl =>
  match retTypM CC (JFClass D0) m with None =>None | Some acid =>
  let newTFR := 
   {| TFRcdecl := cdecl;         TFRmdecl := mdecl;
      TFRXi := thrs_of_md mdecl;
      TFRGamma := loc2env D0 mdecl (JFVLoc (JFLoc n) :: vs);
      TFRfr := fr;               TFRAcid := acid
   |} in Some (h', newTFR :: tfs)
  end end end end end
\end{lstcoq}
while the corresponding one in \coqinl{red} is the following:
\begin{lstcoq}
|(* mthd *)
 (Ctx[[JFInvoke (JFVLoc (JFLoc n)) m vs]]_None) :: Cc =>
   let D0 := getClassName h n in  
     getInvokeBody CC D0 n m vs h Ctx Cc
\end{lstcoq}
In spite of the fact that most of the functionality of \coqinl{red}
for method call is hidden in \coqinl{getInvokeBody}, it is clear
that much more information must be gathered and inspected in
\coqinl{typed_red}.

We proved the correspondence of the two relations in the following two
theorems. The first of them says (soundness) that a well defined step
in the Church version of the semantics implies a corresponding well
defined step in the untyped version: both resulting heaps must agree
and the type erasure \coqinl{FSofTFS} of the resulting typed frame
stack is the resulting untyped frame stack.
\begin{lstcoq}
Theorem fs_from_tfs_after_tred:
  forall h tfs fs h' tfs' res, FSofTFS tfs = fs ->
    typed_red (h, tfs) = Some (h',tfs') ->
    red CC (h,fs) = res ->  res = Some (h', FSofTFS tfs').
\end{lstcoq}
Surprisingly enough the proof in the direction from the typed version
to the untyped one did not require any additional
well-formedness conditions. It was the proof in the opposite direction
(completeness) that used it heavily.
\begin{lstcoq}
Theorem tfs_exists_for_fs_after_red:
  forall tfs h fs h' fs' tfsres,  Well_formed_program CC ->
    type_correct_heap CC h ->  DerivableTFS CC h tfs ->
    FSofTFS tfs = fs ->  well_formed_framestack fs ->
    red CC (h,fs) = Some (h',fs') ->
    typed_red (h, tfs) = tfsres ->
    exists h'' tfs',  Some (h'', tfs') = tfsres /\
      FSofTFS tfs' = fs' /\ h'' = h'.
\end{lstcoq}
Note that apart from typability constraint \coqinl{DerivableTFS CC h tfs}, 
we assume here that the program is well formed %
(\coqinl{Well_formed_program CC}), the heap is consistent with the
program %
(\coqinl{type_correct_heap CC h}), and that the stack used in the
untyped reduction is well formed %
(\coqinl{well_formed_framestack fs}).
\draft{A swoją drogą czy well_formed_framestack nie wynika z DerivableTFS?}


\section{Proving experiments and improvements}
\label{sec:experiments}

\expandafter\ifx\csname SourceFile\endcsname\relax\else\SourceFile{experiments.ltx}\fi
In our proving effort we experimented with a number of techniques to
make the proofs shorter and more readable without sacrificing the easy
to follow optical correspondence with the paper and pencil version. In
some cases the benefits of the experiments were clear from the start
(or from the moment we decided that some improvement is needed) but we
decided to evaluate the impact of given improvements on our proofs
to have a tangible evidence of that.

\paragraph{Format of reduction definitions.}
Our Coq definition of \coqinl{red}, a fragment of which is shown in Section~\ref{sec:semantics},
directly corresponds to the one in Fig.~\ref{fig:semantics}.
Unfortunately, these natural expressions are internally transformed by
Coq into a nested series of matches over particular datatypes in the following order: frame list, context list, context expression, redex expression, execution mode. In particular, even if only a few rules depend on the context (e.g. \srul{letgo} or exception dispatching rules), the matching on context list and last context expression is done systematically and therefore a dozen rules (like \srul{letin}) which do not depend on the context have to be repeated 3 times: 
once for the empty context, and twice for non-empty context: for \coqinl{JFCtxLet} and \coqinl{JFCtxTry}, respectively, as the last context node.
This also means that in proofs over
\coqinl{red}, one would have to handle these repeated rules 
several times. To prevent this, we introduced the
definition \coqinl{red2} where the matching on the
frame stack data structure is done more carefully: first on the access mode $A$ and redex expression $E$ and only if $E$ is a variable, we match on the context to get down to particular exception handling rules (if $A$ is not null), or \srul{mthdret}, \srul{letgo} or \srul{ctchnrml} (if $A$ is null). The definition \coqinl{red2} is a few lines longer and the semantic rules are reordered compared to \coqinl{red}, but the difference in internal representation (obtained by switching off all pretty-printing of matching) is important: 163 vs 476 lines (compared to about 100 lines for each of the definitions in the source files).  
The two reduction definitions can be automatically proven to be equal for
all inputs. In the end we have a close to paper definition
\coqinl{red} that can be switched to
\coqinl{red2} in the proofs and therefore the cases
need not to be repeated. We follow the same approach for our typed
semantics. Again, we have a~natural definition of
\coqinl{typed_red} and the
reordered one called
\coqinl{typed_red2}.

The benefits of these reorderings can be seen in
Fig.~\ref{fig:results} by comparing ``non-duplication'' proofs with
previous ones (S2 with S1 and C2 with C1). For the soundness
proof the gain in script
size was over 50\% and in time about 80\%. For the completeness, the gain in size and
proof generation time was small (about 10\%) but the gain in
proof-checking time (the time of the final \coqinl{Qed}) was also
large (about 40\%). The difference between the two directions
is
caused by the fact that the proofs of particular subcases in the
proof of completeness are significantly longer than ones in the proof of soundness.
Still, the number of cases is always smaller, which explains why the
typechecking took significantly less time even if proof generation time
(which includes searching by \coqinl{auto}) was comparable.

\begin{figure}[t]
  \centering
  \begin{tabular}{|p{210pt}|r|r|r|}
    \hline
    \multicolumn{1}{|p{200pt}|}{\centering ~\linebreak Proof} & 
    \multicolumn{1}{p{30pt}|}{\centering ~\linebreak Size\linebreak in lines} & 
    \multicolumn{1}{p{30pt}|}{\centering Proof generation time} & 
    \multicolumn{1}{p{35pt}|}{\centering Proof\-checking time}
    \\
    \hline
    Soundness non-systematic \hfill (S1)& 
    312 &
    1.636 &
    0.978
    \\
    \hline
    Soundness non-systematic, non-duplication  \hfill (S2)& 
    149 &
    0.429 &
    0.192
    \\
    \hline
    Soundness systematic, automatic  \hfill (S3)& 
    20 &
    0.602 &
    0.264
    \\
    \hline
    Completeness non-systematic \hfill (C1)& 
    498 &
    3.390 &
    1.300
    \\
    \hline
    Completeness non-systematic, non-duplication \hfill (C2)& 
    457 &
    2.927 &
    0.777
    \\
    \hline
    Completeness systematic, non-duplication \hfill (C3)& 
    327 &
    1.798 &
    0.662
    \\
    \hline
    Completeness systematic, automatic blind \hfill (C4)& 
    111 &
    65.177 &
    16.828
    \\
    \hline
    Completeness systematic, well chosen \hfill (C5)& 
    77 &
    4.197 &
    1.534
    \\
    \hline
  \end{tabular}
  
  \caption{Results of experiments. Times are averages from 10 runs on
    the Coq version 8.7.1 on a Lenovo X240 laptop with 8GiB RAM and
    Intel i5-4300U CPU at 1.90GHz running Linux Fedora 26 with kernel
    version 4.16.7.}
  \label{fig:results}
\end{figure}

\paragraph{Comment the cases.}
Both
our main properties relate typed and untyped semantics and therefore both
theorems have premises of the form ``typed semantics gives this'' and ``untyped
semantics gives that'', which both develop into huge nested match expressions.
These premises can be simplified and split into (many many) subproofs
using the \coqinl{destruct} tactic and we developed an
Ltac tactic that does that automatically. While some of the
resulting simple goals are easy to discharge automatically, other
require a manual proof. After some experiments and adjustments we
restricted the number of these \emph{manual} cases to 12 in one lemma
and to 2 in the other. The main difficulty of the process we observed
was that it took us a lot of time 
to understand in which particular case we are left for the
manual proof. To help with this, using a few definitions and
notations, we introduced mock \emph{comment} premises. These comments
are systematically created by our automatic tactics using
\coqinl{destruct} with \coqinl{eqn} argument. In the end, a
typical proof situation to analyse looked like this 
(with about 60 more premises above):
\begin{lstcoq}
d11 : ### A0 = None #
d12 : ### l = JFLoc n0 #
d13 : ### find (TFRGamma e) (JFVLoc (JFLoc n0)) = None #
----------------------------------------------------------------
exists (h'' : Heap) (tfs' : TFSsupport),
  Some (h'', tfs') = None /\ FSofTFS tfs' = fs' /\ h'' = h'
\end{lstcoq}

Although \coqinl{destruct} with \coqinl{eqn}
make these facts present in the list of hypotheses anyway, they are
scattered throughout the list, interspersed with variable declarations
and it is unclear where they came from. Therefore having them together
gathered in one place helps us to realise where we are in the proof in one look.

\draft{ When automatic case analysis is employed, different
  strategies of discharging the cases can be used. We discuss the
  advantages and disadvantages of two approaches to case analysis. In
  the first one, we destruct all the available case distinctions and
  discharge cases where no longer case analysis is possible. In the
  second one, we destruct case distinctions in only one definition and
  apply its results in all the remaining ones. It turns out that
  although the first approach is more general, the second one results
  in shorter proofs.
}

\paragraph{Destruct and discharge.}
Another aspect of working with multiple large definitions by cases is
the strategy used to destruct and discharge goals. If one wants to
automate the process, the easiest approach consists in splitting all
definitions into simple cases by destructing expressions that block 
match constructions, and in the end try to discharge
the remaining goals. When applying this strategy blindly it may happen that the same
expression is destructed many times. 



More precisely, if the destructed expression is a variable, all its
occurrences are replaced by fresh constructor terms (in all subcases), 
so the original variable can never
reappear as a match blocker. However, if one destructs a more complex expression, even though all its occurrences are also replaced by constructors,   
the same expression can be \emph{re-formed}
later in the proof process and it needs to be handled again.
Consider a proof situation where one premise contains a subexpression
\coqinl{match get_sth None with ...} and another one
\coqinl{match get_sth opt with ...} If one destructs first the expression \coqinl{get_sth None} and
later the variable \coqinl{opt}, the expression \coqinl{get_sth None}
reappears (in one of the cases). If it is re-destructed without care,
one can be
confronted with a (false) situation where the case from the second
destruct of \coqinl{get_sth None} is different than the one from
the first destruct. Such situations are sometimes unprovable, but one
can easily prevent them using \coqinl{destruct} with
\coqinl{eqn}. This guarantees that the first choice for
\coqinl{get_sth None} remains in the list of hypotheses, so if the
second choice is different, the context becomes clearly inconsistent,
which can usually be proved by \coqinl{congruence}. Nevertheless,
avoiding multiple destruction of the same expression can significantly 
decrease the number of goals that need to be handled.

In order to do that, before destructing an expression one can first
check the context for some equations concerning it. If the given 
expression has already been destructed, the context must contain an
equation recording the past choice of constructor and it is enough to
reuse that information.

In our development we did precise measurements of the efficiency of
this technique for the proof of completeness. It 
permitted to reduce the number of cases from over a 1000 to 58. Even
though in both versions of the proofs most cases are discharged
completely automatically, the reduction gives dramatic gain in time
(and memory space as well), from minutes to seconds (compare C4 to C5
in Fig.~\ref{fig:results}). The time obtained with the automatic
destruction tactic with early discharge (C5), although significantly
larger than the proof made by systematic manual selection of terms to
destruct (C3), remain in the same comfortable time segment of a small
few seconds, while shortening the proof scrips several times.

It has to be noted that searching context for equations about a
particular expression does not come for free, as one has to search
through all hypotheses using Ltac goal matching. Therefore doing it
systematically in an automatic case analysis is only beneficial if the
probability of repeating an expression is high, which is the case in
our situation.






\section{Related works}
\label{sec:related-works}

There is a significant number of formalisations of semantics for
programming languages so we give here only a rough picture of the
landscape.

The most notable line of research here is done by teams that work on
semantics of the C programming language. One of the most notable
projects here is development of C compiler CompCert in Coq
\cite{2006-Leroy-compcert}. The semantics of current version of
CompCert is primarily expressed in the small-step fashion, but it is
also augmented with a big-step counterpart. These efforts are
complemented by works of other teams (see e.g.\
\cite{KrebbersW15,StewartBCA15}).

Another take on the C semantics was proposed in the context of LLVM
\cite{ZhaoNMZ2012}. The basic non-deterministic small-step semantics
serves to describe the LLVM behaviour. This semantics can be used to
prove the correctness of compilation transformations that operate
locally on instructions. This semantics is further refined into a
deterministic one, which makes it possible to execute pieces of code
and compare results with the actual LLVM platform. There are two more
big-step semantics that are used to prove the correctness of
transformations in case they operate on bigger pieces of code
(functions or program blocks).

Also Maude rewriting system was used to formalise mechanically the
semantics of C \cite{EllisonR2012}. An extension, called K-Maude, was
used for this purpose, which made it possible to define semantics,
which was small-step in spirit, although it heavily used various
structural enhancements of K-Maude.

Formalisations of C compilation process rely on some form of
formalisation of the target low-level language. Typical
formalisations of low-level languages start from small-step semantics
(see e.g.\
\cite{AspertiRCT11,Atkey07,FoxM10,KleinN06,Pichardie06}) and 
only then build big-step versions. This approach makes
it possible to extract \cite{Atkey07,LochbihlerB11} from the proofs an
interpreter of the target language which has a structure compatible
with the imagined process of step-wise computation.

There are formalisations of compiling process for languages other
than~C.  
CakeML is a dialect of ML for which a verified implementation of
compiler was proposed \cite{HupelN18,KumarMNO2014}. The development is
based upon a big-step operational semantics on which the compiler
correctness proof is developed. This is very natural here since
structural operational semantics is the semantics format of choice in
Standard ML \cite{MilnerTHM97}. However, the authors use a small-step
semantics for expressions to define divergence and to make a
type-soundness proof.

In the context of Java-like languages  formalisations
occurred mainly to prove soundness of program analyses. Interestingly
enough, Strni{\v s}a et al \cite{Strnisa10,Strnisa07} presented a
formalisation of a Java-like module system in Isabelle/HOL which
served to identify issues with the existing system as well as to
highlight important design decisions. Even though module semantics are
usually given in big-step form, this formalisation is done in
small-step fashion. This is due to the fact that the goal of the
formalisation was not to give a formal proof of some module system
property, but to have a format in which everything should be expressed
both formally and precisely to gain the confidence in the
understanding of the system.

Recently a number of formalisations appeared
\cite{ArmstrongGS14,ArmstrongGS16,Pous13} that take as their basis
the Kleene algebra with tests introduced by Kozen \cite{Kozen97}. These
works exploit the fact that the high level of abstraction present in
the algebra makes it possible to focus separately on expression forms
that occur in many languages and develop a~general framework to deal
with them and only then instantiate it in the context of a particular
language and property of interest. 

An attractive proposal, similar to the Kleene algebra
with tests, is the generic formalisation of program execution in
small-step fashion by Dinsdale at al
\cite{Dinsdale-YoungBGPY13}. The main motivation for the small-step
format there is that it enables the possibility to express the interleaving
semantics for multiple threads (this advantage was also
observed e.g.\ by Amani et al \cite{AmaniABLRT17}). Additionally, the
system separates the flow of control from actions that manipulate the
state. This facilitates verification of soundness for expressive
type systems in Java-like languages \cite{GordonPPBD12} even in the
context of multithreading. However, the available control flow
expressions make it difficult to handle exceptions and method call
stack so it is difficult to express properties such as that a method
is an extensional function.

In the mentioned above work of Gordon et al \cite{GordonPPBD12} the
method call stack is actually introduced implicitly. This is necessary
to get the soundness argument through since the system, similarly to
ours, requires management of heap regions at the entry and exit from
methods. In order to make the management possible, the authors
introduce an additional non-standard kind of expression
$\mathsf{Bind}$ that binds formal method parameters to the actual ones
and guards the subexpression in which execution of the method body is
performed. In this way the authors obtain implicitly the necessary
functionality of method call stack.

\section{Conclusions}
\label{sec:conclusions}

We presented a Coq formalisation of the frame stack based small-step
operational semantics of \Jafun, a small Java-like language with
classes and an effectual type system that makes it possible to
delineate a notion of a compound value and classify certain methods as
extensional functions. The total size of the whole
formalisation is currently over 16900 lines of proof scripts and
definitions.


For our type system, we defined a notion of typed frame stacks akin to
Church-style expressions in $\lambda$-calculi and proved the
equivalence of the original reduction and the reduction on the
typed stacks. Such a proof turned out to be non-trivial in case of our
system since it required 589 lines of a proof script.

Subsequently we studied different methods this proof could be done to
observe the impact of different approaches on proof construction
efforts and on time of their checking. We measured for example that 
aggressive prevention of repeating destruction of the same expression 
can lead to a tenfold reduction in the proof-checking time.

\bibliographystyle{plain}
\bibliography{functionality-short} 

\end{document}
